# Recent Advances in Real-Time Models for UWB Transmission Systems

**Pierluigi Poggiolini and Yanchao Jiang**
*DET, Politecnico di Torino, C.so Duca Abruzzi 24, 10129, Torino, Italy*
*pierluigi.poggiolini@polito.it*

**Abstract:** Ultrafast accurate physical layer models are essential for designing, optimizing and managing ultrawideband optical transmission systems. We present a closed-form GN/EGN model based on a recent analytical breakthrough, improving reliability, accuracy and generality. © 2025 The Author(s)

## 1. Introduction

Ultrawideband (UWB) optical transmission systems are a promising technology for increasing the capacity of communication networks. Systems using C+L bands are being widely deployed. Research is exploring adding further bands, such as S and E, and others [1]. However, UWB systems are affected by very significant inter-channel Raman scattering (ISRS) which together with the deterioration of most propagation parameters towards higher optical frequencies, makes the design of UWB systems challenging. Careful optimization of launch power and other system parameters is then required. In addition, UWB systems greatly benefit from backward-pumped Raman amplification, but this in turn requires the optimization of the number, frequency and power of the pumps [2].

Optimization is carried out by means of iterative algorithms that require ultra-fast system performance assessment. Non-linear-interference (NLI) estimation is a key component of performance assessment and typically relies on GN/EGN-type models. Since solving the related integrals numerically is out of the question, these models are implemented as closed-form approximate formulas (CFMs). Over the last few years, mainly two groups have pursued the derivation of CFMs, one from University College London (UCL, see [3], [4]), and one from Politecnico di Torino (PoliTo) in collaboration with CISCO (see [5], [6]). Other groups have started looking into similar investigations too.

This paper reports on a new approach at handling the core integrals of the GN model to obtain a CFM. It consists of expressing the spatial power-profile (SPP) of each channel along a span as a polynomial. Once this is done, some of the core GN model integrals admit closed-form near-exact solutions, involving much less drastic approximations than previous CFMs. We call the newly derived result "PCFM", where "P" stands for "polynomial".

PCFM can model UWB systems with ISRS, forward and backward Raman amplification, short spans, lumped loss and other SPPs features. It enjoys better accuracy and reliability than previous CFMs, because it does not hinge on either neglecting intra-span NLI coherence or resorting to infinite series expansions, or other problematic approximations. In the following we first summarize the PCFM derivation and its main formulas. We then show accuracy tests and examples of UWB system results in challenging configurations. PCFM is still work-in-progress: we provide at the end some hints on what directions the development of the approach is taking.

## 2. The Polynomial Closed-Form Model

To derive a CFM from the GN-model integral, some approximations are necessary: (a) channel spectra are rectangular; (b) the power spectral density (PSD) of NLI is computed only at the center frequency of each channel, then NLI is assumed flat over each channel for the purpose of GSNR calculation; (c) SCI and XCI are retained, while MCI is neglected (see [7] for acronyms); (d) fiber loss and dispersion are frequency-dependent, but they are assumed constant over each channel, at the value they assume at each channel's center frequency. All current CFMs, to the best of our knowledge, make all these assumptions. (PCFM could eventually remove (b) and (c), but this is the subject of ongoing investigation.) Then, the PSD of NLI over the "channel under test" (CUT), across a single fiber span, is:

$$G_{\text{CUT}}^{\text{NLI}} = \frac{16}{27} p_{\text{CUT}}^{\text{end}} \cdot \left[ \gamma_{\text{CUT}}^2 P_{\text{CUT}}^3 R_{\text{CUT}}^{-3} K^{\text{SCI}} + 2 P_{\text{CUT}} R_{\text{CUT}}^{-1} \sum_{n_{\text{ch}} \neq \text{CUT}, n_{\text{ch}}=1}^{N_{\text{ch}}} \gamma_{n_{\text{ch}}}^2 P_{n_{\text{ch}}}^2 R_{n_{\text{ch}}}^{-2} K_{n_{\text{ch}}}^{\text{XCI}} \right] \quad (1)$$

where $N_{\text{ch}}$ is the number of WDM channels, $P_{\text{CUT}}$ and $P_{n_{\text{ch}}}$ are the launch power of the CUT and of the $n_{\text{ch}}$-th channel, $R_{\text{CUT}}$ and $R_{n_{\text{ch}}}$ are the respective symbol rates, $\gamma_{\text{CUT}}$ is the SCI nonlinearity coefficient, $\gamma_{n_{\text{ch}}}$ is the XCI nonlinearity coefficient of the $n_{\text{ch}}$-th channel onto the CUT [5]. The factor $p_{\text{CUT}}^{\text{end}} = P_{\text{CUT}}^{\text{end}}/P_{\text{CUT}}$ is the ratio between the power of the CUT at the end of the span and the power of the CUT at launch. $K^{\text{SCI}}$ and $K_{n_{\text{ch}}}^{\text{XCI}}$ are the "core integrals" of the GN-model and constitute the true hurdle to achieving a CFM. Assuming, with no loss of generality, that the origin of all frequencies has been shifted to the center of the CUT, their expressions are:

$$K^{\text{SCI}} = \int_{-R_{\text{CUT}}/2}^{R_{\text{CUT}}/2} \int_{-R_{\text{CUT}}/2}^{R_{\text{CUT}}/2} \left| \int_0^{L_{\text{span}}} p_{\text{CUT}}(z) e^{j4\pi^2 f_1 f_2 \cdot \tilde{\beta}_{2,\text{CUT}} \cdot z} dz \right|^2 df_2 df_1 \qquad K_{n_{\text{ch}}}^{\text{XCI}} = \int_{-R_{\text{CUT}}/2}^{R_{\text{CUT}}/2} \int_{f_{n_{\text{ch}}} - R_{n_{\text{ch}}}/2}^{f_{n_{\text{ch}}} + R_{n_{\text{ch}}}/2} \left| \int_0^{L_{\text{span}}} p_{n_{\text{ch}}}(z) e^{j4\pi^2 f_1 f_2 \cdot \tilde{\beta}_{2,n_{\text{ch}}} \cdot z} dz \right|^2 df_2 df_1 \qquad (2)$$

where $f_{n_{\text{ch}}}$ is the center frequency of the $n_{\text{ch}}$-th channel and $\tilde{\beta}_{2,\text{CUT}}$ and $\tilde{\beta}_{2,n_{\text{ch}}}$ are effective dispersion for the CUT and the $n_{\text{ch}}$-th channel [5]. These integrals use the further approximation of enlarging the exact lozenge-shape integration domain [7] to a rectangular shape inscribing it. This approximation too is performed by all CFMs. The SPPs of the CUT and of the $n_{\text{ch}}$-th channel, $p_{\text{CUT}}(z)$ and $p_{n_{\text{ch}}}(z)$, are normalized to their launch powers: $p_{\text{CUT}}(z) = P_{\text{CUT}}(z)/P_{\text{CUT}}$, $p_{n_{\text{ch}}}(z) = P_{n_{\text{ch}}}(z)/P_{n_{\text{ch}}}$. In early GN-model formulas, the SPPs were expressed by means of fiber loss $\alpha$ as: $p_{n_{\text{ch}}}(z) = \exp(-2\alpha \cdot z)$. Later, frequency-dependence of loss, ISRS and Raman amplification have been introduced, by making $\alpha$ a function of the channel index and of the spatial variable $z$ so that a SPP becomes:

$$p_{n_{\text{ch}}}(z) = \exp\left(-2\int_0^z \alpha_{n_{\text{ch}}}(z) dz\right) \qquad (3)$$

To obtain CFMs, analytical expressions have been used for $\alpha_{n_{\text{ch}}}(z)$ that permit to solve the core integrals in closed form. For instance, $\alpha_{n_{\text{ch}}}(z) = \alpha_{0,n_{\text{ch}}} + \alpha_{1,n_{\text{ch}}} \exp(-\sigma_{n_{\text{ch}}} z)$ was initially proposed in [8], [5] to account for ISRS and forward-Raman amplification. A more complex but similar expression was used to account for backward Raman amplification as well [6]. Different formulas for $\alpha_{n_{\text{ch}}}(z)$ were proposed in [3], [4] to achieve the same purpose. The resulting CFMs, however, require further approximations which are problematic. One example is the need to use a series expansion for certain terms to obtain analytical integration, creating however fractions whose denominators can go to zero. While such singularities are technically removable, they require attention and can generate loss of accuracy. Another example is the approximation needed to deal with backward Raman amplification, which consists of breaking up a span into the start section and the end section. This separation implies the neglect of *coherence* between the NLI produced at the start and at the end of the fiber, that can be substantial. To avoid these problems and acquire more flexibility in representing diverse systems, *we propose to express the normalized SPPs by means of a polynomial*:

$$p_{n_{\text{ch}}}(z) = \sum_{n=0}^{N_p} p_{n_{\text{ch}},n} \cdot z^n \qquad (4)$$

This expression is then inserted into the core integrals Eqs.(2). *Quite remarkably*, the cascaded integrals Eqs.(2) generate *exact closed forms* with no need of any further approximation. A particularly simple solution is obtained for $K_{n_{\text{ch}}}^{\text{XCI}}$ if the single additional approximation of extending the limits of the outermost integral in $f_1$ to $\pm\infty$ is made:

$$K_{n_{\text{ch}}}^{\text{XCI}} = \frac{L_{\text{span}}}{2\pi |\tilde{\beta}_{2,n_{\text{ch}}}|} \left| \log\left(\frac{f_{n_{\text{ch}}} + R_{n_{\text{ch}}}/2}{f_{n_{\text{ch}}} - R_{n_{\text{ch}}}/2}\right) \right| \cdot \left( 2\sum_{n=1}^{N_p} p_{n_{\text{ch}},n} L_{\text{span}}^n \sum_{k=0}^{n-1} \frac{p_{n_{\text{ch}},k} L_{\text{span}}^k}{n+k+1} + \sum_{n=0}^{N_p} \frac{L_{\text{span}}^{2n} p_{n_{\text{ch}},n}^2}{2n+1} \right) \qquad (5)$$

The limit extension for $f_1$ has negligible effects on accuracy as long as $\beta_2$ (ps$^2$/km) $\cdot$ [$R$ (GBaud)]$^2$ > $3 \cdot 10^3$. For $K^{\text{SCI}}$ the same integration limit extension is not possible. Closed-form formulas are still found for each set value of $N_p$ but no compact expression like Eq.(5), with $N_p$ as a parameter, has been derived yet (research is ongoing). The closed-form results for $K^{\text{SCI}}$ for $N_p$=0 and 1 are, as an example:

$$R = R_{\text{CUT}}, \quad L = L_{\text{span}}, \quad x = \pi^2 \tilde{\beta}_{2,\text{CUT}} R^2 L, \quad S = \sin(x), \quad C = \cos(x), \quad \text{SI} = \text{SinIntegral}(x), \quad H = {}_2F_3\left(\left\{\frac{1}{2}, \frac{1}{2}\right\}, \left\{\frac{3}{2}, \frac{3}{2}, \frac{3}{2}\right\}, -\frac{1}{4}[x]^2\right)$$

$$K_{N_p=0}^{\text{SCI}} = 2R^2 L^2 p_0^2 H + (1-C)(2p_0^2)/(\pi^4 \tilde{\beta}_{2,\text{CUT}}^2 R^2) - (2Lp_0^2 \text{SI})/(\pi^2 \tilde{\beta}_{2,\text{CUT}})$$

$$K_{N_p=1}^{\text{SCI}} = \left[ 2p_1^2 + 9R^4 \tilde{\beta}_{2,\text{CUT}}^2 (2p_0^2 + 2Lp_0 p_1 + L^2 p_1^2) \pi^4 - 2(p_1^2 + \pi^4 \tilde{\beta}_{2,\text{CUT}}^2 R^4 (9p_0^2 + 9Lp_0 p_1 + 4L^2 p_1^2)) C + 6R^8 \tilde{\beta}_{2,\text{CUT}}^4 L^2 \cdot \right.$$

$$\left. \cdot (3p_0^2 + 3Lp_0 p_1 + L^2 p_1^2) \pi^8 H - 2R^2 \tilde{\beta}_{2,\text{CUT}} Lp_1^2 \pi^2 S - 2R^6 \tilde{\beta}_{2,\text{CUT}}^3 L(9p_0^2 + 9Lp_0 p_1 + 4L^2 p_1^2) \pi^6 \text{SI} \right] / \left( 9R^6 \tilde{\beta}_{2,\text{CUT}}^4 \pi^8 \right)$$

We have found the $K^{\text{SCI}}$ formulas for $N_p$ up to 9 [9], but it is of interest to limit the polynomial degree as much as possible since the number of $K^{\text{SCI}}$ terms increases roughly as $2^{\wedge}N_p$. In Fig.1, we show the polynomial fitting Eq. (4) with $N_p$=3,5,9 of an S-band channel in a C+L+S 150-channel system with ISRS and backward Raman. The channel has high loss at the start due to ISRS and high gain at the end due to Raman, representing a challenge for the polynomial fitting. The fit with $N_p$=9 appears excellent, but not so for $N_p$=5 or 3. Quite remarkably, though, the NLI error incurred using $N_p$=5 or even 3 is minimal. In Fig.2 we show the violin plots of the error calculated between PCFM and our reference numerically-integrated full EGN model, over the 150 channels of the mentioned C+L+S system after the first span (100km of SMF). The plots show small error in general, and no need to go beyond $N_p$=5. Note that the +0.5dB systematic bias in Fig.2(a) is due to the assumption that NLI is locally white.

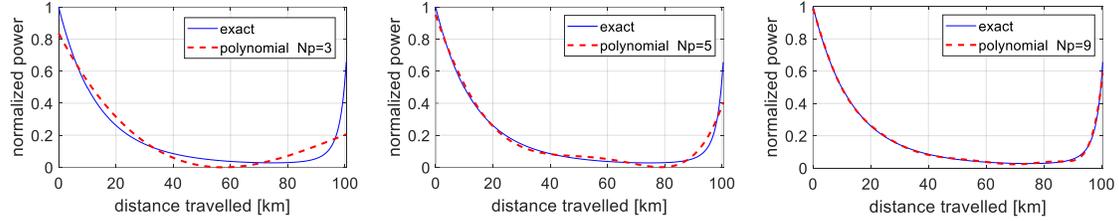

**Fig.1: polynomial rendering of the spatial-power-profile $p_{n_{ch}}(z)$ of a mid-S-band channel in a 100km Raman-amplified C+L+S system.**

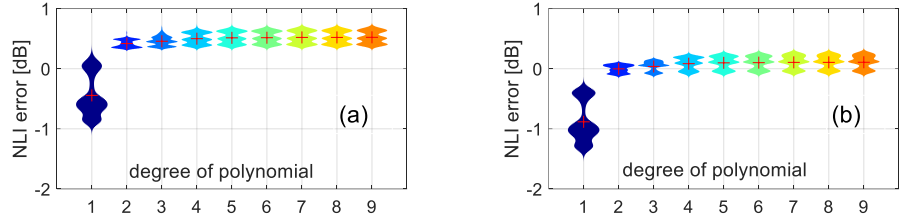

**Fig.2: Violin plots of the NLI estimation error for the single-span C+L+S system described in the text. (a): PCFM vs. EGN-model; (b): same, but the correction $\rho$ [10] is used. The abscissa is the degree of the polynomial used to represent channel power profiles in Eq.(4).**

PCFM could remove this assumption but this is left for future research. However, once the PCFM is combined with the machine-learning correction coefficient $\rho$ introduced in [10], the bias disappears (Fig.2(b)). The quite small residual error attests to the reliability of the PCFM. The PCFM also allows to deal with situations that cannot be tackled by other CFMs. For instance, lumped loss. Fig.3 shows for the same system of Fig.2 the presence of 2dB lumped loss at 5 km. The resulting NLI GSNR is shown in Fig.3. The correspondence with the markers found using the reference EGN model is excellent with and without lumped loss. Note that Eq.(1) accounts for NLI in one span. For multi-span systems the PCFM NLI contribution of each span are combined as done for previous CFMs [10].

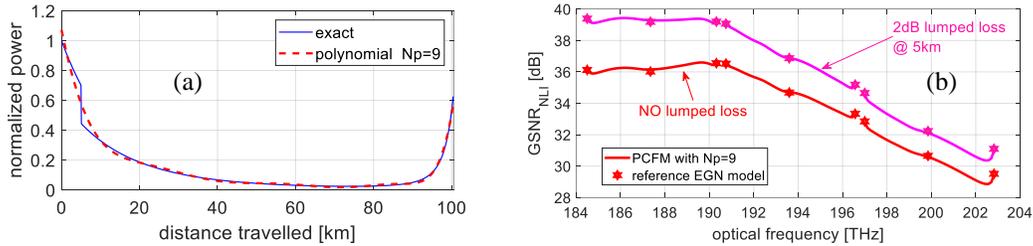

**Fig.3: (a) power profile of the same channel as in Fig.1, with added 2dB lumped loss at 5km. (b) GSNR (NLI only) for the 150-channel C+L+S system with Raman amplification described in the text. Markers are the reference EGN model, with and without the lumped loss.**

### 3. Comments and conclusion

We have presented a new closed-form GN/EGN model, called Polynomial-CFM or PCFM, based on an analytical breakthrough which leads to improved CFM reliability, accuracy and generality. We have shown results and examples of the accuracy and generality of the PCFM, for UWB systems with ISRS and Raman amplification. Research is ongoing, since the approach promises to allow tackling more general system configurations, that were difficult to deal with by previous CFM, such as lumped loss, subcarrier transmission, very low dispersion and others.

This work was partially supported by: Cisco Systems RISEONE research contract; the PhotoNext Center of Politecnico di Torino; the European Union under the Italian National Recovery and Resilience Plan (NRRP) of NextGenerationEU, partnership on "Telecommunications of the Future" (PE00000001 - program "RESTART").